# Close packing density and fracture strength of adsorbed polydisperse particle layers

Robert D. Groot and Simeon D. Stoyanov
Unilever Research Vlaardingen,
P.O. Box 114, 3130 AC Vlaardingen, The Netherlands

The close packing density of log-normal and bimodal distributed, surface-adsorbed particles or discs in 2D is studied by numerical simulation. For small spread in particle size, the system orders in a polycrystalline structure of hexagonal domains. The domain size and the packing density both decrease as the spread in particle size is increased up to 10.5±0.5%. From this point onwards the system becomes amorphous, and the close packing density increases again with spread in particle size. We argue that the polycrystalline and amorphous regions are separated by a Kosterlitz-Thouless-type phase transition. In the amorphous region we find the close packing density to vary proportional to the logarithm of the friction factor, or cooling rate. We also studied the fracture behaviour of surface layers of sintered particles. Fracture strength increases with spread in particle size, but the brittleness of the layers shows a minimum at the polycrystalline-amorphous transition. We further show that mixing distributions of big and small particles generally leads to weaker and more brittle layers, even though the close packing density is higher than for either of the particle types. We point out applications to foam stability by the Pickering mechanism.

## 1. Introduction

The close packed state of spherical particles in 3D has a long history of study[1] because of its many practical applications. E.g. it has been used as model for sandpile stability, liquids, glasses and to locate the point where suspension viscosity diverges. Recent work led to a simple expression for the close packing density in 3D for any size distribution;[2] in particular for bimodal distributions small particles can form a 'rattler-phase' in between of big particles that form a stress-bearing network. It has further been established that the dense random packing density in 3D depends upon the friction between the particles.[2,3] Much less attention has been payed to the 2D case, and many questions are still open.

Experiments in 2D with bubble rafts[4] and $n$ monodisperse discs of diameter $d$ per unit area on a moving air table[5] show a maximum packing of $y_0 = (\pi/4)nd^2 \sim 0.84$ and a bit larger for irregular particles or a tilted table.[6] One of the aims of this paper is to study the close packing density of polydisperse discs in 2D. The effect of polydispersity on freezing in 2D was studied by Pronk and Frenkel,[7] and more recently by Fingerle and Herminghaus.[8] However, the role of friction in the 2D case is not clearly defined in theory or simulation.[8,9,10] This is another point of concern.

The close packed state in 2D is important for understanding the stabilisation of air bubbles in foams by colloidal particles, the Pickering mechanism. Bubbles are often stabilised by proteins, surfactants or fat, but this route is limited by Ostwald ripening, a process that leads to coarsening of the foam and eventually complete loss of air from a product.[11,12,13] In the last ten years however, attention is drawn again towards the use of colloidal particles (10-1000 nm) to stabilize emulsions or foams.[14] Stabilisation of foams by particles is beneficial for a number on reasons. First of all, particles generally have a much higher adsorption energy than proteins or surfactants.[15] Furthermore, because of their relatively large size they form a steric layer around bubbles.

Particle-particle interactions are shown to play a crucial role in the stability of a foam.[15,16] These interactions may lead to fractal surface structures or to dense solid layers, depending on the interaction strength and surface coverage, as pointed out by Groot and Stoyanov.[17] The surface pressure is often many orders of magnitude larger than expected from a simple hard sphere equation of state, as explaned recently.[17] This is very important for the stabilization mechanism.

In addition to this, particles also form highly rigid and elastic layers that prevent Ostwald ripening. The surface elastic modulus is crucial for stability. If the bubbles are surrounded by a strong elastic layer, big bubbles cannot inflate so that disproportionation is halted. The foam becomes unstable when the surface layer is ruptured. In the disproportionation process the small bubbles shrink, while the big bubbles expand, leading to coarsening. As these two processes are linked, it is often argued that stopping only bubble shrinkage is sufficient to stop the disproportionation, and that using irreversibly adsorbed colloidal particles is an effective way to achieve this, because such particles have a finite maximum packing at the interface.[14–16]

In reality, however, particle–covered bubble surfaces can also wrinkle or buckle, leading to volume loss at constant surface area. In addition, initial surface defects early after foam formation allow for surface rearrangements and small bubble shrinkage; and gas pockets that arise after pouring foam into a container form a source of gas. These factors drive the growth of large bubbles, especially in early stages, leading to significant changes in bubble size distribution and eventually to foam collapse. Therefore, to stop the disproportionation process efficiently, both bubble shrinkage and bubble growth should be halted. The most effective interfaces therefore provide sufficient elasticity towards both expansion and compression. As discussed before, to stop shrinkage repulsive core interactions are sufficient, but to stop bubble expansion also attractive (sticky) interactions are needed. Ultimately, the strength of a solid surface layer is determined by its fracture properties.

If we could manipulate surface fracture behaviour, we could therefore design foam stability. The important questions for this are, can we manipulate the fracture strength of the interfacial layer by particle size and size distribution? And can we influence how brittle the surface layer will be? Are surface layers weaker when the particles form big hexagonal domains? If so, how polydisperse should the particles be to prevent domain formation? To address these questions we first investigate dense random packing of model surface layers, as this forms a natural point of reference, and subsequently use these systems to study their fracture behaviour. In section 2 the simulation model is described, results on surface ordering are presented in section 3, close packing is described in section 4 and these results are used to interpret fracture strength in section 5. We summarize conclusions in section 6.

## 2. Simulation model and methods

### 2.1 Conservative forces

The simplest model to represent colloidal particles is to use elastic spheres with short-range attraction, this is the sticky elastic sphere model. A detailed account for monodisperse particles is given elsewhere.[17,18] This model is generalised here for particles of arbitrary size. The linear elastic repulsion between two soft overlapping spheres of radii $a_i$ and $a_j$ that are fused at dihedral angle $\psi$ takes the form

$$F_{ij}^{Rep} = \tfrac{\pi}{3} E a_i a_j \sin(\psi)\left(1 - \frac{r}{D_{ij}}\right) \qquad 1$$

where $r$ is the distance between particle centres, $D_{ij}$ is the distance of closest approach given dihedral angle $\psi$, and $E$ is the linear elastic modulus, see Figure 1.

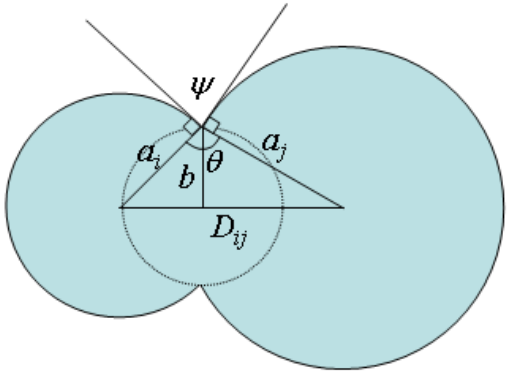

**Figure 1** Fused spheres at dihedral angle $\psi$. The sphere radii are $a_i$ and $a_j$, equilibrium distance between centres is $D_{ij}$, and the central circle of radius $b$ indicates the contact zone where most deformation takes place.

At separation distances further away than $D_{ij}$ the force should become attractive because particles at contact form hydrogen bonds, or have a strong hydrophobic interaction. A simple form to represent this, is to assume a parabola force,

$$F^{Att} = \frac{4\varepsilon_{ij}}{\delta_{ij}^2}(1 - r/D_{ij})(1 + \delta_{ij} - r/D_{ij}) \qquad 2$$

The parameter $\delta_{ij}$ is the range of the attractive interaction relative to the mean diameter, and $\varepsilon_{ij}$ is the force minimum. We now make three requirements for the force field:
1. the repulsive force follows linear elasticity for $r < D_{ij}$;
2. the derivative of the force is continuous in $r = D_{ij}$;
3. the reversible work to fracture a bond is the net surface energy of the contact area.

The first condition leads to the force given in Eq 1. The other conditions lead to two simple relations between the force amplitude and the force range. From these conditions and the linear elastic law the force range $\delta_{ij}$ and amplitude $\varepsilon_{ij}$ follow as

$$\delta_{ij} = 6\sqrt{\gamma b(1-\cos(\psi/2))/ED_{ij}^2}\;;$$
$$\varepsilon_{ij} = \tfrac{1}{2}\pi b^{3/2}\sqrt{E\gamma(1-\cos(\psi/2))} \qquad 3$$

Here, $b$ is the radius of the neck, which is given by $b = a_i a_j \sin(\psi)/D_{ij}$ and $D_{ij}$ is the distance of closest approach, which is given by $D_{ij} = (a_i^2 + a_j^2 + 2a_i a_j \cos\psi)^{1/2}$.

The factor $1-\cos(\psi/2)$ appears in Eq 3 because the energy to fracture a bond is the energy difference $G = 2\pi\gamma b^2 - \pi\gamma_g b^2$, where $\gamma$ is the surface energy of the particle-solution interface, and $\gamma_g$ is the surface energy of the grain boundary between two particles. In equilibrium these two are related via $\gamma_g = 2\gamma\cos(\psi/2)$.

It is convenient to redefine the above relations for the limit $\psi \to 0$ to simulate particles with additive core interactions. In this limit the effective modulus is $E_0 = E\psi$. For small dihedral angle the mean diameter is $D_{ij} = a_i + a_j$ and the neck radius is $b = R_{ij}\psi/2$, where $R_{ij}$ is the harmonic mean radius,

$$R_{ij} = \frac{2a_i a_j}{a_i + a_j} \qquad 4$$

Furthermore, the adhesion energy $G_{ij} = 2\pi\gamma b^2(1-\cos(\psi/2)) \approx 2\pi\gamma R_{ij}^2\psi^4/32$, hence we define the surface tension

$$\gamma_0 = \gamma\psi^4 \qquad 5$$

Thus, we arrive at the following force range and force amplitude in the limit $\psi \to 0$:

$$\delta_{ij} = \frac{3}{2D_{ij}}\sqrt{\frac{\gamma_0 R_{ij}}{E_0}}\;;\quad \varepsilon_{ij} = \frac{\pi}{16}R_{ij}^{3/2}\sqrt{E_0\gamma_0} = \frac{\pi}{24}\delta_{ij}R_{ij}D_{ij}E_0 \qquad 6$$

In the present simulations we assume finite and constant values for $E_0$ and $\gamma_0$. The previous derivation serves to derive the right scaling relation of forces between particles of different sizes.

Apart from a radial force between particles, fused solid particles also interact by bending and shear forces. These can be estimated by assuming all deformation to take place in a contact zone of radius $b$, and using linear elasticity theory to calculate the elastic response to a given affine deformation.

The torque between particles that are bent over an angle α with respect to the contact point is thus derived as

$$\tau \approx -\frac{4Eb^3}{3\pi}\alpha + O(\alpha^3) \qquad 7$$

If we now substitute $E = E_0/\psi$ and $b = R_{ij}\psi/2$, we find that the torque is proportional to $\tau \propto \psi^2$, hence the bending rigidity vanishes for small dihedral angle. The shear force, on the other hand does not vanish. The shear force in the particle centres is given by

$$\mathbf{F}_i^s = -\mathbf{F}_j^s = -\frac{\pi E b D_{ij}}{9 r_{ij}}\mathbf{u}_{ij}^\perp \qquad 8$$

where $\mathbf{u}_{ij}^\perp$ is the perpendicular displacement of the particles relative to the point of contact. The product $Eb = E_0 R_{ij}$ remains finite when $\psi \to 0$. In general, the shear force also leads to torques on the particles, such that the total angular momentum of the two particles is conserved. In the limit $\psi \to 0$ these are given by

$$\begin{aligned}\boldsymbol{\tau}_i &= a_i \mathbf{r}_{ij} \times \mathbf{F}_i^s / D_{ij} \; ; \\ \boldsymbol{\tau}_j &= a_j \mathbf{r}_{ij} \times \mathbf{F}_i^s / D_{ij}\end{aligned} \qquad 9$$

where $\mathbf{r}_{ij} = \mathbf{r}_j - \mathbf{r}_i$.

Shear forces are included if a solid network is simulated. In that case a neighbour list is used. Particles are neighbours if they are within interaction distance. Attractive and shear interactions are applied to listed pairs with an intact bond. The shear interaction is defined with respect to the point of closest contact at the time the bond was formed. When a bond is broken only the repulsive interaction is used (Eq 1).

## 2.2 Friction, Brownian forces, and particle masses

The precise scaling of the hydrodynamic interaction between two spheres was discussed for the 3D case by Farr and Groot.[2] This work showed that the squeeze mode of the lubrication force between neighbouring particles takes the form

$$\mathbf{F}_{ij}^{sq} = -R_{ij}\gamma\, f(h/R_{ij})(\mathbf{v}_{ij}\cdot\mathbf{e}_{ij})\mathbf{e}_{ij} \qquad 10$$

where $h = r - D_{ij}$ is the gap width between the particle surfaces, $\mathbf{e}_{ij} = \mathbf{r}_{ij}/|\mathbf{r}_{ij}|$ is a unit vector pointing from particle $i$ to particle $j$, and $R_{ij}$ is again the harmonic mean radius. In fact the scaling function $f(x)$ for small arguments diverges as $f(x) \sim 1/x$, but there is a large body of evidence showing that correct long-range inertial hydrodynamics is generated even if the divergent lubrication force between particles is replaced by a finite distance-dependent friction.[19,20] Thus, we use the friction function[2]

$$\mathbf{F}_{ij}^{frict} = -R_{ij}\gamma\left(1 - h/R_{ij}\right)^2 \theta(1 - h/R_{ij})\,(\mathbf{v}_{ij}\cdot\mathbf{e}_{ij})\mathbf{e}_{ij} \qquad 11$$

where $\theta(x)$ is the Heaviside step function. This captures the right physics regarding the scaling of the range and strength of the viscous interaction with particle sizes.

It is a somewhat arbitrary choice to truncate the friction factor at a distance of a particle radius $R_{ij}$ but in a dense suspension hydrodynamic screening will shield off all interaction from larger distances anyway. To compensate for the energy loss by friction, Brownian noise can be introduced, see e.g. Groot and Stoyanov.[18] In this case a radial random force is introduced, given by

$$\mathbf{F}_{ij}^R = \sqrt{2R_{ij}\gamma k T}\left(1 - h/R_{ij}\right)\theta(1 - h/R_{ij})\,\zeta_{ij}(t)\,\mathbf{e}_{ij}/\sqrt{\delta t} \qquad 12$$

where $\zeta_{ij}(t)$ is a random number of unit variance, and $\delta t$ is the time step taken. Note that the amplitude and distance-dependence of the noise function follow from the fluctuation-dissipation theorem.[18]

It has been shown by Farr and Groot[2] that for this system the close packing density depends on the particle size, friction factor and friction range. The reason is that the close packing density is kinetically determined, and therefore depends upon friction.[3] This implies firstly that, to compare different particle sizes, the range of the friction function must be scaled relative to the particle size. Secondly, in a system of elastic particles two time scales are pertinent: a elastic oscillation time and a drag relaxation time. To have a uniquely defined close packing density, the ratio of these two time scales must be fixed. This ratio is given by $t_{el}/t_d = 2\pi\gamma(D/Em)^{1/2}$, where $D$ is the particle diameter and $m$ is its mass.

The close packing density can only be independent of particle size if we maintain a constant ratio between these two time scales, hence the friction should follow $\gamma \propto (m/D)^{1/2}$. Since the friction factor is independent of particle mass and size (it is only related to the fluid viscosity), a well-defined (size-independent) close packing density is obtained if and only if we choose the particle mass proportional to $m \propto D$. This choice has been made here. The predicted scaling was checked by simulation and holds exactly.[2] In reality the influence of friction is small, as shown below.

To specify the model in physical units, we choose the maximum particle diameter as unit of length, $D^* = 1$; and the corresponding mass is used as unit of mass $m^* = 1$. For this diameter we arbitrarily choose the slope of the force in Eq 1 as $\pi E_0/12 = 1000$. This fixes length, mass and time scale. Once the relative force range $\delta_{ij}$ for particles of diameter $D_{ij} = 1$ is chosen (generally $\delta_{ij} = 0.03$ for $D_{ij} = 1$), the force range and amplitude follow for all particle sizes from Eq 6.

## 2.3 Polydispersity

Particles deposited on or adsorbed to a surface are generally not monodisperse, but usually follow a log-normal distribution. To simulated this we generate a distribution of particle sizes with modal radius $a_0$. The distribution is generated by multiplying the modal radius $a_0$ for each particle by $\exp(\sigma\zeta_i)$, where $\zeta_i$ is a standard Gaussian random number for particle $i$, with spread $\sigma$.

For any distribution of particles the polydispersity is usually expressed via the polydispersity index PDI, defined as

$$PDI = \frac{M_w}{M_n} = \frac{\sum_i M_i^2}{\sum_i M_i} \cdot \frac{N}{\sum_i M_i} = \frac{\langle M^2 \rangle}{\langle M \rangle^2} \quad 13$$

If we have a single peak log-normal particle size distribution, the probability density to find a particle with radius within a small interval around $a$ is given by

$$dP(a) = \frac{1}{\sigma\sqrt{2\pi}} \exp\left(-\frac{[\ln(a/a_0)]^2}{2\sigma^2}\right) d\ln(a/a_0) \quad 14$$

By straightforward integration we find the $n^{\text{th}}$ moment of the distribution as $\langle a^n \rangle = a_0^n \exp(n^2\sigma^2/2)$. The polydispersity index thus follows as PDI = $\langle a^6 \rangle/\langle a^3 \rangle^2$ = $\exp(9\sigma^2)$. Reversely, the spread corresponding to a distribution of given polydispersity is obtained as $\sigma = (\ln(\text{PDI}))^{1/2}/3$.

## 3. Surface ordering

Monodisperse spherical particles on a surface at high density generally form hexagonal domains. If particles assemble on an air bubble, and if these domains are large, the size of the domain may interfere with the strength of the surface layer. Generally, we may expect that the layer is weaker when the domains are larger, as this implies that there are long grain boundaries where the layer may rupture under stress. We expect therefore that, to obtain a strong surface layer, it is desirable to have small domains. Therefore we first study how the domain size is influenced by the particle size distribution.

When particles are large, Brownian motion can be neglected. In that case we can simulate a T = 0 system by leaving out the random noise altogether. Starting with a random conformation, friction will slowly take out kinetic energy from the system until all particles come to a standstill. Domains of different orientation are separated by grain boundaries where the local hexagonal structure is broken. The mismatch between particles at these grain boundaries increases the free energy of the system. As long as there is a significant amount of energy to gain, smaller domains will be subsumed the bigger ones, leading to domain growth via an Oswald ripening process. This is illustrated in Figure 2.

The system depicted in Figure 2 consists of 20,000 particles of radius 0.3832, confined to an area of size $L_x \times L_y$ = 102×102. The particles interact with a repulsive force $\pi E_0/12$ = 1000, i.e. the slope of the force in Eq 1 would be 1000 for particles of diameter $D = 1$, and has been scaled accordingly for particles of smaller diameter. Further to the repulsive interaction, the beads have a friction interaction as given in Eq 11, with friction factor $\gamma = 1$.

To measure the local hexagonal order, we define the following vectorial order parameter for every particle $i$:

$$\mathbf{s}_i = \left(\frac{1}{6}\sum_j \cos(6\varphi_{ij}), \frac{1}{6}\sum_j \sin(6\varphi_{ij})\right) \quad 15$$

where $\varphi_{ij}$ is the angle between any pair of neighbours and the $x$-axis. Neighbours are defined here as particles within friction interaction distance, i.e. a centre-to-centre distance less than $r$ = $D_{ij}+R_{ij}$. For six neighbours the order parameter takes its maximum value if the neighbours are ordered regularly, in which case $|\mathbf{s}| = 1$. If particles have fewer than six neighbours the order parameter is reduced proportionally; for any regular array of more than six neighbours $|\mathbf{s}| = 0$. The particle order parameter is mapped onto a colour via $(r,g,b) = (½+½s_1, ½–¼s_1+½s_2, ½–¼s_1–½s_2)$. Particles with a vanishing order parameter are mapped onto grey; and when $|\mathbf{s}| = 1$ the colour is saturated.

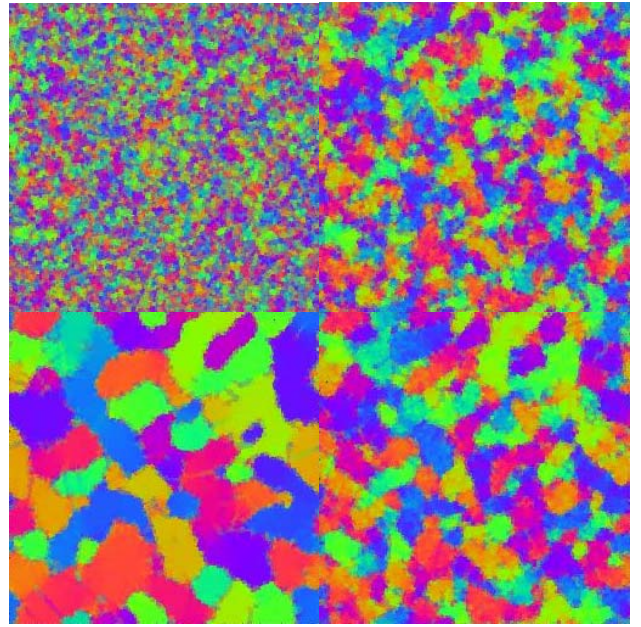

**Figure 2** Time series of ordering in 2D surface, time t = 2, 3, 4 and 10 (clockwise, starting from top left) Different colours indicate a different orientation of local hexagonal ordering.

To analyse the domain size, the surface is divided into a grid (of 80×80 cells) and the order parameter is pre-averaged over all particles within a cell. Thus, the order parameter is mapped onto 2D 'spin' vectors on a regular square lattice. Next, the spin-spin correlation function is averaged over all lattice sites:

$$C(r) = \tfrac{1}{2}<\mathbf{s}(x,y)\cdot\mathbf{s}(x+r,y)>_{xy} + \tfrac{1}{2}<\mathbf{s}(x,y)\cdot\mathbf{s}(x,y+r)>_{xy} \quad 16$$

Since periodic boundary conditions are applied the maximum value of $r$ in Eq 16 is half the system size. The spin-spin correlation function is shown in Figure 3 for a number of evolution times. This figure shows that the growth of domains has virtually come to a standstill after $t = 25$; the correlation function at $t = 100$ has barely increased.

Each of the correlation functions in Figure 3 represents the order in a single conformation. Because the system has a finite size, the correlation function fluctuates around zero at large distances. To obtain a statistical mean, an average of 25 independent systems was taken, that were all evolved over 10 time units. This correlation function is shown in Figure 4, together with the correlation functions shown in Figure 3 that were obtained for different evolution times, scaled to a time-dependent correlation length. This firstly shows that for an infinitely large system the correlation function decays continuously, and secondly it shows that within the noise the correlation function shows scaling in time. From the statistically averaged correlation at $t = 10$ an accurate fit function was derived that may serve to define the correlation length. To a large degree of accuracy the correlation function follows

$$C(r,t) = C(0,t) \exp[-(r/\xi(t))^{1.5}] \qquad 17$$

The observed scaling behaviour implies that the size distribution and the shape of the domains remains constant in time as the ageing process continues, up to a scaling factor for the size.

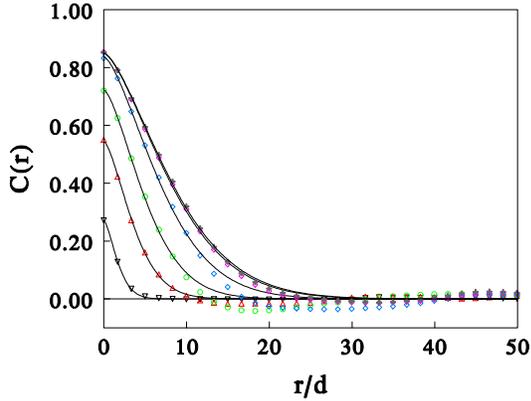

**Figure 3** Spin-spin correlation of a single system at time t = 2 (black), 3 (red), 4 (green), 10 (blue), 25 (magenta) and t = 100 (grey).

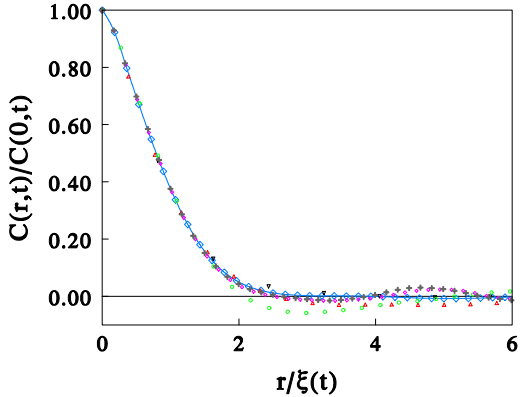

**Figure 4** Spin-spin correlation of the systems shown in Figure 3, scaled to the time-dependent correlation length. Blue curve and symbols give an average over 25 systems of age $t = 10$.

The time dependence of the correlation length $\xi(t)$ has not been analysed thoroughly, but qualitatively $\xi$ increases with time up to a maximum value in what appears to be a stretched-exponential behaviour. The importance of the scaling behaviour given in Eq 17 is that it provides a reliable way of defining the domain size. Thus we can investigate how the domain size depends upon the particle size distribution.

Systems with log-normal distributed particle radii were prepared (see section 2.3) but radii above $a > 0.5$ were downsized to $a = 0.5$. All systems contained 20,000 particles that were spread over an area of 102×102. In all cases the attractive interaction was set at $\varepsilon = 0$ and the friction was set at $\gamma = 1$. No noise was added. All systems were evolved over 5000 steps or more with $\delta t = 0.01$. Statistical accuracy was increased by averaging over 5 different samples for all systems except for $\sigma = 2.5\%$, for which 12 independent starting conformations were generated. The correlation length of fully aged systems ($t = 50$) is shown in Figure 5.

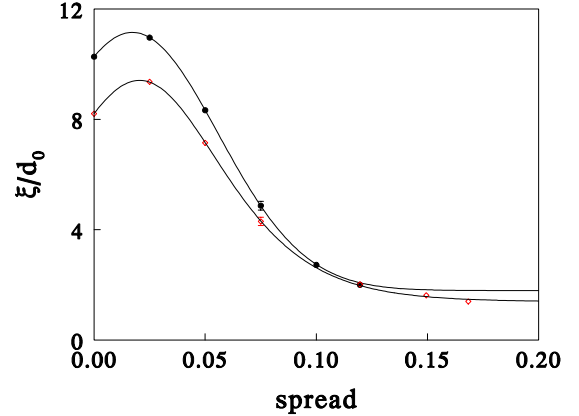

**Figure 5** Correlation length at constant area 102×102 (black dots), and at constant surface pressure $\Pi \approx 0.5$ (red dots).

Remarkably, we find a maximum in the correlation length near $\sigma = 1.7\%$ (PDI = 1.0027). We speculated that this is caused by pressure increase with increasing spread. To check this we also simulated at constant surface pressure. These results are shown in Figure 5 by the red dots. Some influence of the surface pressure on the domain size is found, but the maximum in the correlation length remains. Therefore this maximum is *not* an artefact of the constant area. The main result is that the domain size shrinks to one grid point by $\sigma$ = 10-12.5% (PDI = 1.10-1.15).

## 4. Dense random packing in 2D

### 4.1 Unimodal particle size distributions

One aim of this work is to simulate the strength of surface-adsorbed layers of solid particles. To simulate fracture strength, we must prepare solid particle networks at vanishing tension, otherwise the stress-strain curve will start at a non-zero stress. However, particles with a strong short-range attraction typically quench in a non-equilibrium state with many vacancies, and depending on the starting density a state of zero pressure can be obtained for a wide range of densities. Therefore we choose a well-defined state of reference. The close packed state is a good starting point to define the density of vanishing spreading pressure for systems with attraction.

First we return to the set of systems described in section 3. Log-normal particle size distributions are generated, and for each value of the spread in the particle size distribution the area was varied. The surface pressure obtained at T = 0 in all cases shows a nearly linear increase for area fractions $y = (\pi/4) \Sigma d_i^2/A > y_0$, and vanishes for $y < y_0$. The area fraction $y_0$ is the close packing density in 2D. To find this point efficiently, we use a variation of the Lubachevsky-Stillinger method,[2,21] the surface area is increased or decreased to maintain a constant pressure. All systems were prepared at an initial area fraction $y = 0.9$ and evolved over 100 time units to obtain equilibrium. Then a constant surface pressure $\Pi_{ext} = 0.01$ was imposed and the systems were evolved over another 400 time units. In all cases the friction factor was $\gamma = 1$. The results are shown in Figure 6.

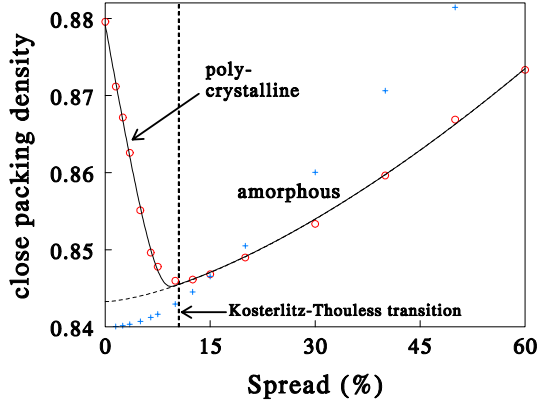

**Figure 6** Dense random packing density as function of the spread in the particle size distribution (red circles). Small blue crosses give the prediction of Farr-Groot theory,[2] fitted to 15% spread and adapted for 2D systems, see Eq 22.

For small spread in particle size the maximum packing density decreases from 88% down to 84.6% at a spread of $\sigma \approx 11\%$. From that point onwards the close packing density increases again with spread. To understand this behaviour we draw an analogy between particles at an interface, and the XY model. We have seen that particles that are not too different in size form hexagonal domains. These domains can be characterised by a spin vector $s_i$ (Eq 15). When a domain is perfectly hexagonal $|s_i| = 1$, but the orientation of the spin vector varies by 360° when the orientation of the domain is rotated by 60°. The energy of the system is clearly minimal when all spins are oriented in the same direction; the energy of a grain boundary corresponds to the mismatch between neighbouring spins. Therefore the system may be described by the Hamiltonian

$$H = J \sum_{ij} (1 - s_i \cdot s_j) \qquad 18$$

where the sum runs over *neighbouring* spins $i$ and $j$. This is the XY model. It is one of the rare models in statistical physics that could be solved exactly.[22,23,24,25] In particular, the model has an infinite order phase transition between an ordered phase at low temperature and a disordered phase at high temperature, which is known as the Kosterlitz-Thouless transition. This transition has been associated with melting in two dimensions.[26,27,28]

Generally, the energy in Eq 18 corresponds to the total length of grain boundary in the system; in the particle system these reduce the maximum packing density. Therefore we have a close analogy between dense random packing in a particle system and the Hamiltonian of the XY model, where spread ($\sigma$) in the particle system corresponds to temperature ($1/J$) in the spin model. Hence, to fit the data of Figure 6 we use the known scaling function of the XY model, which has an essential singularity at the transition point, and we add a regular function to describe the increase of packing at large spread:

$$y_{cp} = y_0 + a\sigma^{3/2} + b\exp\left(\frac{-c}{\sqrt{1-\sigma/\sigma_t}}\right)\theta(1-\sigma/\sigma_t) \qquad 19$$

Using this fit function, the transition point from polycrystalline to amorphous is obtained as $\sigma_t \approx 10.5 \pm 0.5\%$. The extrapolation of the power law curve to zero spread (dashed curve in Figure 6) gives the area fraction as $y_0 = 0.8433 \pm 0.0003$. Finally, the coefficient $a$ is obtained as $a = 6.5 \pm 0.1 \times 10^{-5}$ if the spread is given in percent. The notion of a Kosterlitz-Thouless type transition for polydisperse discs was also suggested by Santen and Krauth.[29]

The interpretation of this result is that for a distribution with a spread less than 11% the system forms recognizable domains that decrease in size when the spread is increased, because bigger particles frustrate regular packing. The smaller the domains, the larger the total area occupied by grain boundaries. Since grain boundaries raise the free energy, the surface pressure is increased and hence the close packing density is decreased. This is consistent with the polycrystalline state observed for $\sigma < 10\%$ (Figure 2).

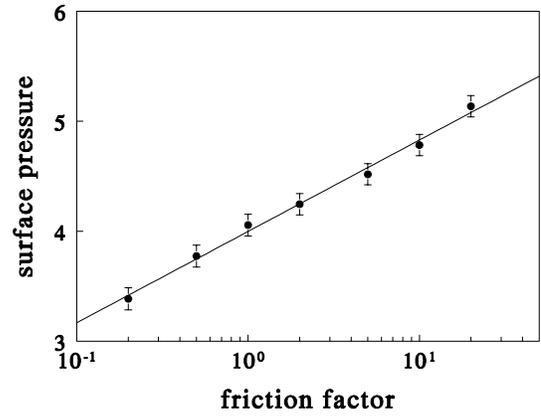

**Figure 7** Mean pressure at $T = 0$ as function of friction factor, for area fraction $y = 0.8574$ and for spread $\sigma = 17.5\%$.

The data points for spread $\sigma > 12\%$ all fall on a power law that increases with spread. This suggests that around $\sigma \sim 11\%$ the system shows a transition from polycrystalline to amorphous. Indeed, by $\sigma = 10\%$ the correlation length has decreased to about the grid size and remains virtually the same for larger spread. When the spread is increased further, the system has more possibilities to pack small particles in the holes between the big ones; therefore the close packing density increases again with increasing polydispersity, see Figure 6.

Another point to check is how the close packing density in a 2D system depends on the friction factor. For this check we chose the system at σ = 17.5%. The modal radius was $a_0$ = 0.3832; and a size cut-off was imposed for $a > 0.5$. All simulated systems have an area $L_x \times L_y = 106^2$, and the friction factor was varied from γ = 0.2 to γ = 20. The spreading pressure in this series of systems at $t = 50$, averaged over 50 runs, is shown in Figure 7. Over a variation of two decades, we find the spreading pressure at $y = 0.8574$ to increase with the logarithm of friction. The implies that also the close packing density varies with log(friction), since $y_0 \approx y - \Pi/\kappa$, where $\kappa = \partial\Pi/\partial y \approx 364 \pm 15$ at the close packing density. Even though the effect is small in practice (limited to the third decimal place of $y_0$ over the range of frictions considered), this behaviour is quite different from the behaviour in 3D, see Farr and Groot.[2]

### 4.2 Bimodal particle size distributions

From Figure 6 we see that a log-normally distributed particle system forms polycrystalline domains for small spread in particle size (σ < 10%), and an amorphous state for larger spread (σ > 12%). However, in absolute sense the packing density does not increase by much, even for very large spread.

The increasing packing density with spread as shown in Figure 6 suggests that higher densities can be obtained by further increasing the spread in particle size. However, for mono-modal distributions a wider distribution than a 60% spread is not practical, because the neighbour search for large size ratios becomes slow. For the largest spread considered (60%) the size ratio between the biggest and smallest particle in the sample is roughly 1:50. To increase the spread, bimodal distributions are therefore considered. To define the simulated systems, we first introduce the area fraction of big particles as

$$w = \frac{N_{big} a_{big}^2}{N_{big} a_{big}^2 + N_{small} a_{small}^2} \qquad 20$$

This is the fraction of area that is covered by big particles, relative to the total covered area.

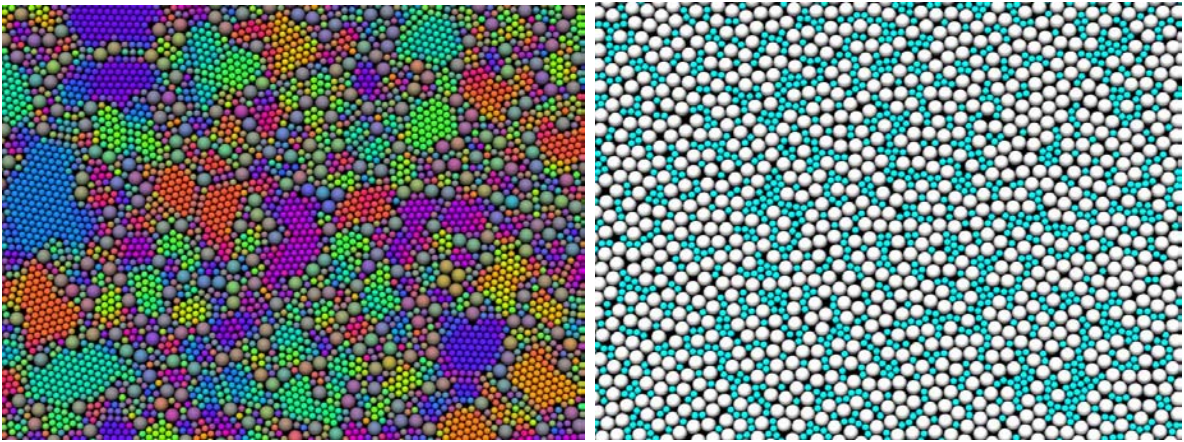

**Figure 8** Samples of binary packing for size ratio $D_{big}/D_{small}$ = 2. Left: low fraction of big particles ($w$ = 0.25), right: ($w$ = 0.75). Colours of left figure follow the orientational order parameter, Eq 15.

First, a binary system of size ratio 1:2 was simulated. The larger particle has a modal diameter of $D_{big} = 1$, and the smaller one has diameter $D_{small} = 0.5$. Qualitatively, we find a maximum in packing, as in 3D, but crystallization plays an extra role. We observe crystalline domains of small particles between large particles for a low fraction of big particles, and separate crystalline domains of small and of big particles for a high fraction of big particles, see Figure 8. Hence, we have entangled effects of crystallization and packing.

To unravel the role of particle size dependence we need to prevent crystallization. This can be done effectively by turning the binary distribution into a bimodal distribution. We are well in the amorphous phase for a spread of σ = 15% (see Figure 6). Therefore we mix two log-normally distributed populations of 15% spread. Each population would by itself lead to a close packing density $y_0 = 0.8468 \pm 0.0001$. Mixing big and small particles together will allow the small ones to position into the gaps between the big particles, thus leading to a higher area fraction. To allow for the width of the distribution, we take the modal diameter of the fraction of large particles as $D_{big} = 0.6$, and the other fraction has modal diameter $D_{small} = 0.6/R$, where $R = 2, 5$ or 10 is the size ratio.

**Table I** 2D close packing density of bimodal mixtures[a]

| w | R=2 | R=5 | R=10 |
|---|---|---|---|
| 0 | 0.8468 | 0.8468 | 0.8468 |
| 0.25 | 0.8535 | | |
| 0.4 | | 0.8783 | |
| 0.5 | 0.8578 | | 0.8990 |
| 0.6 | | 0.8919 | |
| 0.62 | 0.8588 | | |
| 0.65 | | | 0.9139 |
| 0.7 | | 0.8964 | |
| 0.75 | 0.8576 | | 0.9224 |
| 0.8 | | 0.8966 | 0.9240 |
| 0.85 | | | 0.9184 |
| 0.88 | 0.8541 | | |
| 0.9 | | 0.8861 | 0.9079 |
| 0.95 | | 0.8709 | 0.8833 |
| 1 | 0.8468 | 0.8468 | 0.8468 |

[a]Mixtures of two log-normal distributions are used, each having a spread of 15%; $w$ is the area-averaged fraction of big particles, and $R$ is the size ratio of the two distributions.

The results for the three size ratios are shown in Figure 9 and summarized in Table I. All systems contain 20,000 particles, and in all cases the initial area fraction was $y = 0.9$. With friction factor γ = 1 all systems were evolved over 100 time units ($10^4$ steps) to bring the system to rest. Then constant surface pressure simulations were done (using $\Pi_{ext}$ = 0.01) until the area fraction was stable over five decimal places. The maxima in the random close packing density were obtained by polynomial fits, and are summarized in Table II.

**Table II** Maxima in 2D close packing density[b]

| $R$ | $w_{max}$ | $y_{max}$ |
|---|---|---|
| 1 | 0.5 | 0.8468±0.0001 |
| 2 | 0.624±0.007 | 0.8587±0.0001 |
| 5 | 0.750±0.005 | 0.8975±0.0005 |
| 10 | 0.787±0.003 | 0.9241±0.0002 |
| ∞ | 0.8672±0.0001 | 0.9765±0.0001 |

[b]Mixtures of two log-normal distributions are used, each having a spread of 15%; $w_{max}$ is the area-averaged fraction of big particles at the maximum, and $y_{max}$ is the maximum attainable packing fraction.

The two black dashed curves in Figure 9 present the upper limit for the close packing density for infinite size ratio. The rising curve is obtained by starting with a fully packed surface of small particles, and replacing patches of small particles by a big particle. Here, locally the area fraction $y_0$ = 0.8468 is replaced by 1. The decreasing curve corresponds to a situation where the big particles form a jammed structure (with $y_0$ = 0.8468) and the small ones fill up the holes. For infinite size ratio local structuring effects around the big particles can be neglected, and we find[2]

$$y_{cp} = \max\left\{\frac{y_0}{1-w(1-y_0)}, \frac{y_0}{w}\right\} \quad 21$$

For infinite size ratio the maximum is found for a fraction of big particles $w_{max}$ = $(2-y_0)^{-1}$ = 0.8672, at the value $y_{max}$ = $y_0(2-y_0)$ = 0.9765. The numbers apply to populations that each have a 15% spread around their mean size. The variation of $y_0$ with spread is small, however, so in practice these numbers for other spreads will be very similar.

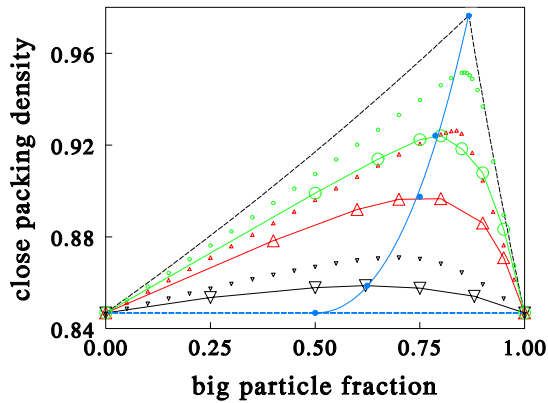

**Figure 9** Dense random packing in 2D for mixtures of two log-normal distributions, each having a spread of 15%. The big particle fraction is defined in Eq 20. The size ratio is given by $D_{big}/D_{small}$ = 2 (black), 5 (red), 10 (green). Blue curve gives position of maxima. Small black, red and green dots give theoretical predictions.

A theory for the close packing density in 3D was recently pubished by Farr and Groot.[2] This theory can be generalised to arbitrary space dimension, and has been adapted by us to 2D systems. For the 2D case the size distribtution of discs is mapped onto a distribution of 1D rods via

$$P_{1D}(L) \propto L \int_L^\infty \frac{P_{2D}(s)}{\sqrt{s^2-L^2}} ds \quad 22$$

which replaces Eq (1) of ref [2]. For $P_{2D}$ we substituted the sum of two log-normal distributions, and we used the free-volume parameter $f$ = 0.2428 in Eq (2) of ref [2] to fit the 2D close packing density at 15% spread. We then use the algoritm given in ref [2] to pack $10^4$ 1D rods. We find that, although the theory is very accurate for 3D systems, it fails in 2D. Some qualitative behaviour shown in Figures 6 and 9 is reproduced, but the theory systematically overestimates the increase of the close packing density for mixed systems.

## 5. Surface layer fracture strength

### 5.1 Unimodal particle size distributions

To test the strength of a particle-covered interface we have to bind the particles together by an interaction force. The particles are modelled by linear elastic springs. As above, the spring constant is chosen as $\pi E_0/12$ = 1000. We can choose the interaction range and force only for one particle size; we take $\delta$ = 0.03, thus for $D$ = 1, $\varepsilon$ follows as $\varepsilon$ = $(\pi E_0/12) \delta/4$ = 7.5. For all other particle sizes the attractive force range and amplitude follow from the scaling relations given in Eq 6.

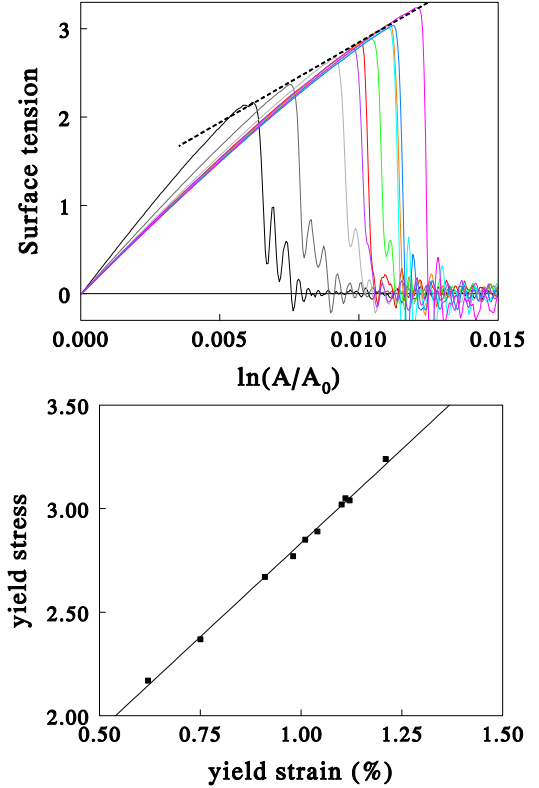

**Figure 10a (top)** Tension-strain curves for log-normal distributions of spread $\sigma$ = 0, 1.5%, 2.5, 3.5, 5, 6.5, 7.5, 10, 12.5 and 15%. Spread $\sigma$ = 0 is black, 1.5 and 2.5 are grey, and the other curves follow spectral order. **b (bottom)** the yield points of these curves.

To prepare a solid surface layer, the dense random packed systems of the previous section are used, but an attractive force of δ = 0.03 and ε = 7.5 for D = 1 is added. This leads to a slight surface pressure drop. Because the spring constant is finite, the systems slightly contract to maintain a constant surface pressure, typically the density increases by 0.3-0.6% if the external surface pressure is maintained at 0.01.

After the surface network is equilibrated to the external spreading pressure Π = 0.01, it is stretched by increasing the surface area by a factor 1.00001 every 20 time steps. This results in an area increase rate $d\ln(A)/dt = 5 \cdot 10^{-5}$ (using time step $\delta t = 0.01$). A typical value for the area increase at failure is around 1%, which is reached in some 200 time units. Since the oscillation time of the system is about 11 time units, the system is allowed to oscillate some 18 times before it breaks. A faster deformation rate would lead to fewer oscillations, hence to a less accurate determination of the failure properties.

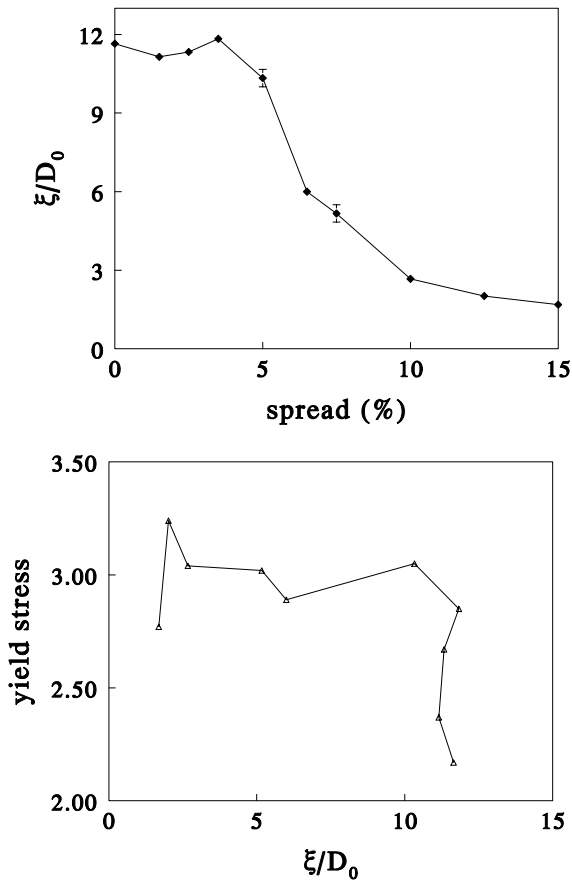

**Figure 11a (top)** Correlation length as function of spread and **b (bottom)** yield stress as function of correlation length.

We start with log-normal particle size distributions of modal particle diameter $D_0 = 0.6$, and vary the spread in the distribution. The stress-strain curves and the yield point for systems with 15% spread or less are shown in Figure 10. The yield point is the position of the maximum of this curve. There are a number of notable results. Firstly, the slope of the curves at vanishing deformation – i.e. the modulus – decreases with increasing spread. This is quite unexpected. However, yield stress and yield strain generally increase with spread, but the order is not completely absolute. When yield stress is plotted against yield strain we find a perfectly linear correlation as shown in Figure 10b.

We hypothesized that surface layers get stronger when the domain size is smaller, because this would imply fewer (weak) grain boundaries in the system. To test this hypothesis the correlation length as defined in Eq. 17 was determined. The remarkable result is that correlation length is constant up to 5% spread, and then it starts to decrease. This is in line with the results of Figure 5. In contrast, the yield stress as function of spread increases with spread for σ < 5%, but levels off at higher spread. Thus, if we correlate yield stress to correlation length (Figure 11b) we obtain a curve with a horizontal branch that corresponds to systems of spread $5 \leq \sigma \leq 12.5\%$, and a vertical branch that corresponds to $0 \leq \sigma \leq 5\%$. Hence there is no obvious relation between yield stress and correlation length. The spread in particle size seems more important than the corresponding correlation length.

So far we have restricted to polycrystalline systems, apart from the highest two spreads. Hence, the behaviour shown in Figures 10 and 11 refers to the left part of Figure 6. Now we will explore systems to the right of the polycrystalline-amorphous transition. To accommodate for the large spread in the size distribution (a maximum size $D = 1$ is used) the modal diameter of these systems was reduced to $D_0 = 0.2$, and the spread was varied from 20 to 60%, in steps of 10%. The stress-strain curves were determined according to the above protocol; the results are shown in Figure 12. Quite surprisingly, for these systems a fit through the yield points is sloping downward, as opposed to the upward trend in Figure 10. Thus, polycrystalline surface layers and amorphous surface layers behave qualitatively different in this respect.

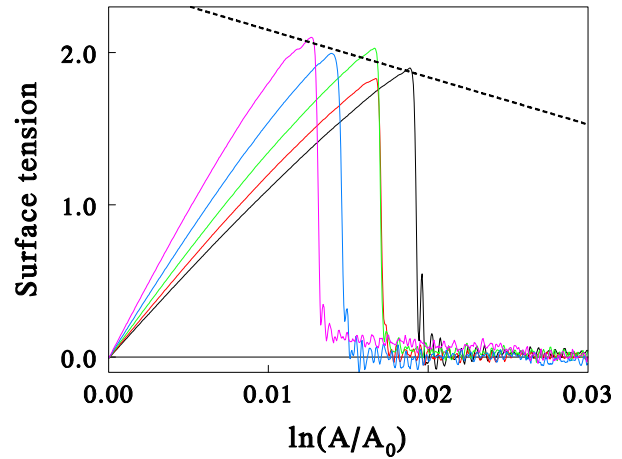

**Figure 12** Stress-strain curves for σ = 20% (black), 30% (red), 40% (green), 50% (blue) and 60% (magenta). The dashed curve is a linear fit through the yield points.

To compare the two data sets quantitatively, we need a scaling relation for the yield point as function of the mean particle diameter. If all systems would have the same number of bonds per particle, we could use the scaling relation for the interaction force. Because the interaction range is only a few percent of the particle diameter this will be a good first order approximation. From the scaling relations in Eq 6 we have $\varepsilon \sim R_{ij}^{3/2}$ and $\delta \sim R_{ij}^{1/2} D_{ij}^{-1}$, where $R_{ij}$ is the mean harmonic radius and $D_{ij}$ is the mean diameter. Both of these are proportional to the modal diameter $D_0$. In two dimensions the yield stress is proportional to the maximum bond force $\varepsilon$ divided by particle diameter, hence $\sigma_y \sim \varepsilon/D_0 \sim D_0^{1/2}$. The yield strain should be proportional to the force range, hence $\gamma_y \sim \delta \sim D_0^{-1/2}$. Concluding, we expect that yield curves of systems of different particle sizes but the same spread fall onto the same scaling curve, which takes the form

$$\sigma D_0^{-\nu} = f\left(D_0^{\nu} \ln(A/A_0)\right) \qquad 23$$

where the scaling exponent should be $\nu = \frac{1}{2}$.

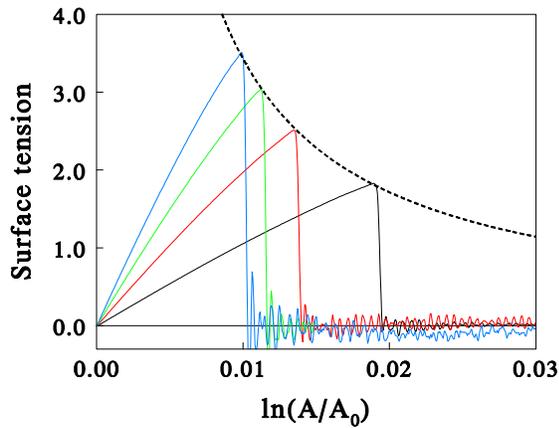

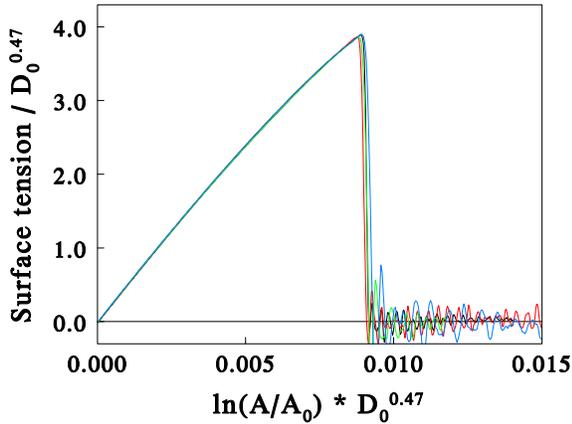

**Figure 13** (top) Stress-strain curves for 10% spread systems of modal diameter $D_0$ = 0.2 (black), 0.4 (red), 0.6 (green), and 0.8 (blue). The lower graph shows the same data, plotted with scaling exponent $\nu$ = 0.47.

To check this scaling relation, four systems of 10% spread in diameter were simulated. The modal particle diameter was varied as $D_0$ = 0.2, 0.4, 0.6 and 0.8. The expected scaling relation appears to hold reasonably well, but if we replace the theoretical exponent $\nu = \frac{1}{2}$ by a semi-empirical number, we find excellent scaling behaviour for $\nu$ = 0.47, as shown in Figure 13. The deviation from $\nu = \frac{1}{2}$ is attributed to the fact that the number of bonds per particle slightly depends on the force range. Because the yield stress is proportional to $\sigma_y \sim D_0^{\nu}$ and the yield strain $\gamma_y$ (= $\ln(A/A_0)$ at the yield point) is proportional to $\gamma_y \sim D_0^{-\nu}$ we have the hyperbolic relation $\sigma_y = 0.034/\gamma_y$, which is shown by the dashed curve in Figure 13. This relation will become important to interpret the behaviour of binary mixtures.

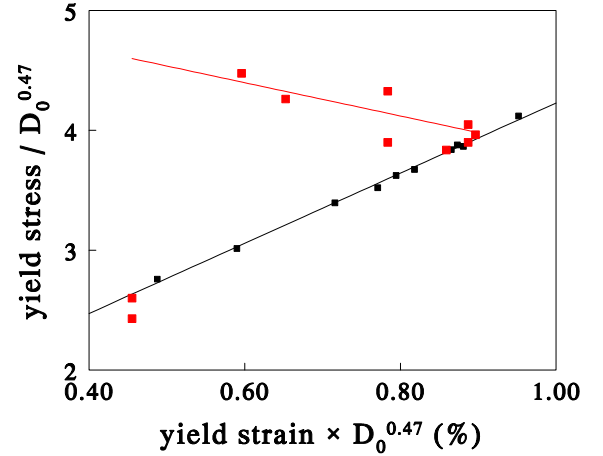

**Figure 14** Yield point for all simulated systems in single scaling graph. The red dots are systems of modal diameter $D_0$ = 0.2, the black dots have modal diameter $D_0$ = 0.6.

This confirms our earlier conclusion that systems of spread $\sigma < 11\%$ behave qualitative different from spread $\sigma > 11\%$. The former systems are polycrystalline, the latter are amorphous. The former are brittle and weak, the latter are brittle but strong. The two phases are most probably separated by an infinite order Kosterlitz-Thouless transition. The least brittle system is found exactly at the transition point.

When we apply this scaling relation to the above simulation results for particles of modal diameters $D_0$ = 0.2 and $D_0$ = 0.6, we may collect all yield data in the same plot, Figure 14. To extend the data set for $D_0$ = 0.2 some extra systems were generated at spread $\sigma$ = 0, 1.5, 5, 10 and 15%. These systems have yield points that follow the correlation obtained for the larger diameter systems.

### 5.2 Bimodal particle size distributions

Finally we study the fracture strength of bimodal layers. Figure 6 shows that, as a system becomes more polydisperse, the close packing density first goes down from 88.0% to 84.6% where the polycrystalline-amorphous transition is reached; then the random packing density goes up again. Nevertheless, a density as high as for monodisperse systems is not reached even for a polydisperse system of spread $\sigma$ = 0.6. One may expect that a higher surface density will lead to higher yield strength. To increase the area fraction, bimodal systems were studied in section 4.2. This indicated that for a size ratio 1:5 an area fraction of 89.7% can be achieved, which exceeds the value for monodisperse particles. Generally, we have seen that bigger particles lead to stronger systems. The obvious question is, can we make stronger layers by mixing two systems of largely different size ratio?

To study this point we used eight systems of size ratio 1:5. Each system was first pressure equilibrated over $10^4$ steps with attractive interaction ε = 7.5 and δ = 0.03. Next, all systems were evolved over 6×10$^4$ steps, with an area increase by 1.00001 after every 20 steps, the same as in the previous section.

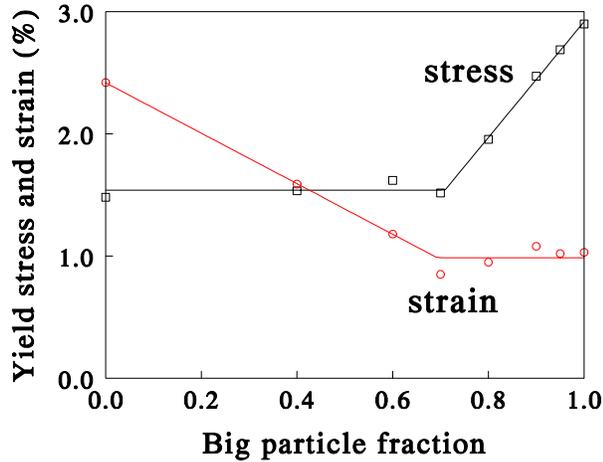

**Figure 15** Yield point for binary systems as function of the area fraction of large particles. The black curve and symbols give the tension at which the surface layer fractures, and the red curve and symbols give the area increase at failure.

The resulting yield stress and strain are shown in Figure 15. Contrary to expectation, a mixture of big and small particles does not lead to reinforcement of the surface layer. Instead the strength of a layer of big particles is undermined by adding small particles. For all compositions with $w<w_{max}$ (≈ 0.7) we find the yield stress of the small particle system and a reduced yield strain; and for all compositions of $w>w_{max}$ we find the yield strain of the big particle system but a reduced yield stress, see Figure 15. A fit of the maximum density (with attractive interaction) gives the maximum as $w_{max}$ = 0.727±0.006. If we fit the stress and strain data in Figure 15 to two kinked lines, we find the cross-over points at $w_{stress}$ = 0.71±0.02 and $w_{strain}$ = 0.69±0.04 respectively. These three values are the same within the error.

The observed behaviour can be explained as follows. For $w<w_{max}$ the big particles are surrounded by a continuum of small ones (like holes in a Swiss cheese). The modulus of a small particle network is smaller than for big particles, hence all deformation is concentrated in the small particle connections. Consequently, the yield stress is that of a small particle network. The *local* deformation within a patch of small particles at failure is given by the yield strain of a small particle network. Since the big particles behave as rigid structures, the area increase relative to the total area is smaller than the local strain within the 'cheese' of small particles.

For $w>w_{max}$ the big particles form a stress-bearing network. This is the 'rattler-phase' where small particles can rattle around in the holes between the big particles.[2] This network will fail when the total strain exceeds the yield strain of a big particle network. Whenever this network fails locally, all stress is loaded upon the small particles that must subsequently fail because the local strain is larger than the global strain. Therefore the yield strain is constant. We speculate that the yield stress in this region is linear in the weight fraction because some of the big-big connections are replaced by big-small connections, which are weaker. As in regular solution theory, the number of big-big connections per big particle increases proportional to the fraction of big particles.

The firm conclusion is that mixing the same type of particles with not too large size difference weakens the surface layer. Strictly speaking we have no information of larger size ratio than 1:5, but the above reasoning indicates that for $w_{max} < w < 1$ the network strength depends on the big particle fraction $w$ because the number of strong bonds per big particle is diluted down if small particles are added. Therefore we may expect the same relation to hold also for bigger size ratios.

## 6. Summary and conclusions

It is known that air bubbles can be stabilised by particles at the water-air interface. The stability depends on the surface elasticity and more in particular on the fracture strength of the surface layers.

To study the scope of manipulating the strength of these layers by particle size, and size distribution, simulation work was done on model systems. It is hypothesized that spherical particles may form domains at the surface, and that the domain size influences the fracture strength, and hence bubble stability. It is further hypothesized that a larger coverage of the surface leads to stronger layers, and that increased coverage can be obtained by mixing big and small particles.

We show that particles indeed form domains at the surface by ordering in hexagonal patterns. We define an order parameter that allows precise measurement of the hexagonal domain size. The size of these domains depends on the spread in the particle size distribution. As the spread increases the domain size is reduced and at spread $\sigma_t$ = 10.5±0.5% the system undergoes a transition from polycrystalline to amorphous. To understand the nature of this transition, an analogy is made with the so-called XY model that has been solved exactly. This suggests that the transition is an infinite order phase transition.

In connection to this we determined the random close packing density of polydisperse hard discs in two dimensions. This shows that up to the transition to the amorphous phase the packing density decreases; from then on the packing density increases again, but even at a 60% spread the packing density is lower than for a monodisperse system. A recent theory that maps the problem of packing polydisperse hard spheres in 3D onto the problem of packing rods in 1D has been modified to describe the 2D case. We show that, although the theory is very accurate for 3D packing, it is only qualitatively correct in the 2D case.

For systems in the polycrystalline phase the fracture strength and strain both increase with the spread in particle size. However, beyond the polycrystalline-amorphous transition the yield strain decreases again with increasing spread, while the yield strength keeps increasing with spread in particle size.

When particle size is increased the model predicts the fracture strength to increase roughly as $\sigma_y \propto D^{0.5}$, while the yield strain decreases as $\gamma_y \propto D^{-0.5}$. Hence, bigger particles make strong but brittle surface layers.

For bimodal distributions where big and small particles are mixed, and where each of these populations has a 15% spread, we find that the close packing density has a maximum as function of area-weighted fraction $w$ of bigger particles. The maximum increases with the size ratio of the two populations and the position $w_{max}$ shifts towards higher fractions of big particles. Systems with smaller fractions of big particles ($w<w_{max}$) form a 'Swiss cheese phase', where the small particles form a continuous network and the big particles act like holes in this 'cheese'. Systems with larger fractions of big particles ($w>w_{max}$) form a 'rattler phase' where small particles rattle around between the big particles.

Mixing different sizes of the same particle type generally weakens the surface layer. In the 'Swiss cheese phase' the small particles form a continuous network. Consequently, the yield stress equals that of a small particle network, but yield strain varies linearly with $w$. In the 'rattler phase' the big particles form a stress-bearing network, but its strength is reduced because the number of big-big connections is diluted down by small particles. Therefore the yield stress here varies linearly with $w$, and the yield strain equals that of a network of the big particles. For particles of unequal type different behaviour may however be observed, this is outside the scope of the present work.